\documentclass[aps, prl, floatfix, nofootinbib, superscriptaddress, twocolumn]{revtex4-1}

\usepackage{latexsym}
\usepackage{amsmath}
\usepackage{amssymb}
\usepackage{amsfonts}
\usepackage{slashed}

\usepackage[mathscr,scaled=1.15]{urwchancal}
\DeclareFontFamily{OT1}{pzc}{}
\DeclareFontShape{OT1}{pzc}{m}{it}%
{<-> s * [1.15] pzcmi7t}{}
\DeclareMathAlphabet{\mathpzc}{OT1}{pzc}{m}{it}

\usepackage{color}

\usepackage{supertabular}
\usepackage{placeins}
\usepackage{enumitem}
\usepackage{epsfig}
\usepackage{graphicx}

\definecolor{purple}{rgb}{0.5,0,0.5}
\definecolor{blue}{rgb}{0.0,0,0.9}
\definecolor{prdblue}{rgb}{0.133,0.118,0.498}
\usepackage[colorlinks=true, pdfstartview=FitV, linkcolor=prdblue, citecolor= prdblue, urlcolor=prdblue]{hyperref}

\begin{document}

\title{$\,$\\[-7ex]\hspace*{\fill}{\normalsize{\sf\emph{Preprint no}. NJU-INP 026/20}}\\[1ex] Resolving the Bethe-Salpeter kernel}

\author{Si-Xue Qin}
\affiliation{Department of Physics, Chongqing University, Chongqing 401331, China.}

\author{Craig D.~Roberts}
\affiliation{School of Physics, Nanjing University, Nanjing, Jiangsu 210093, China}
\affiliation{Institute for Nonperturbative Physics, Nanjing University, Nanjing, Jiangsu 210093, China\\
\hspace*{1.6em} sqin@cqu.edu.cn \hspace*{8.7em}
cdroberts@nju.edu.cn
}

\begin{abstract}
A novel method for constructing a kernel for the meson bound-state problem is described.  It produces a closed-form that is symmetry-consistent (discrete and continuous) with the gap equation defined by any admissible gluon-quark vertex, $\Gamma$.  Applicable even when the diagrammatic content of $\Gamma$ is unknown, the scheme can foster new synergies between continuum and lattice approaches to strong interactions.  The framework is illustrated by showing that the presence of a dressed-quark anomalous magnetic moment in $\Gamma$, an emergent feature of strong interactions, can remedy many defects of widely used meson bound-state kernels, including the mass splittings between vector and axial-vector mesons and the level ordering of pseudoscalar and vector meson radial excitations.\\[1ex]
\end{abstract}


\date{2021 May 26}

\maketitle

\noindent\emph{1:\;Introduction}.\,---\,
%
Spectroscopy has long been crucial in searching for an understanding of Nature's fundamental forces.  The strong interaction spectrum began to demand attention following discovery of $\pi$-mesons \cite{Lattes:1947mw, Bjorklund:1950zz, Panofsky:1950gj}.  In quantum mechanics models \cite{GellMann:1962xb, GellMann:1964nj, Zweig:1981pd}, these pions are bound-states of a constituent-quark, $Q$, and constituent-antiquark, $\bar Q$, with two-body angular momentum $L=0$,  aligned spins ($S=0$), and principal quantum number $n=1$ (radial ground state):
they are $1\,^1\!S_0$ states.

The $\pi$-mesons' spin-flip excitations, the $1\,^3\!S_1$ $\rho$-mesons, were discovered a decade later \cite{Erwin:1961ny}; then the orbital excitations ($1\,^1\!P_1$ $b_1$-mesons) \cite{Abolins:1963zz}; and the orbital excitations of the $\rho$-mesons ($1\,^3\!P_1$ $a_1$-mesons) \cite{Gavillet:1977kx, Daum:1980ay}.
A $1\,^3\!P_0$ companion, the $\sigma$-meson, wherein the $Q\bar Q$ pair possesses $L=1$, $S=1$, $J=L+S=0$, was long controversial \cite{Basdevant:1972uu}; but a picture of a broad scalar resonance has recently become accepted \cite{Pelaez:2015qba}.
Plainly, it is not spectroscopy unless the $n=2$ excitations of these states are located.  Candidates for the $2\,^1\!S_0$, $2\,^3\!S_1$ and $2\,^3\!P_1$ states have been identified; but the $2\,^1\!P_1$ state is \emph{missing} and the complexity of the $1\,^3\!P_0$ resonance suggests that it will be difficult to identify and understand a $2\,^3\!P_0$ state.
There are gaps at $n=3$; and little is known about $n\geq 4$ mesons \cite{Zyla:2020}.

Theoretical frameworks must be developed and employed which can translate this sparse empirical information into statements about quantum chromodynamics (QCD).  In particular, one must expose the character of emergent phenomena in QCD which lead to the existence of mesons, explain their nature, and produce the ordering observed in the spectrum \cite{Roberts:2020udq, Roberts:2020hiw}.

Chiral symmetry is dynamically broken in QCD.  Pions emerge as the associated Nambu-Goldstone bosons \cite{Nambu:1960tm, Goldstone:1961eq}.  Dynamical chiral symmetry breaking (DCSB) is a pivotal corollary of emergent hadronic mass (EHM) \cite{Brodsky:2015aia, Brodsky:2020vco, Barabanov:2020jvn}.  It is characterised by a momentum-dependent quark mass-function \cite{Bhagwat:2003vw, Bowman:2005vx, Fischer:2005nf, Bhagwat:2006tu}, which is large at infrared scales, even when the Higgs coupling to light-quarks vanishes, and key to understanding $>98$\% of visible mass in the Universe.  An insightful spectrum calculation should elucidate consequences of this aspect of EHM.
%
%

\smallskip

\noindent\emph{2:\;Symmetry constraints on the scattering kernel}.\,---\,%
%
The properties of any colour-singlet system constituted from a valence-quark and a valence-antiquark can be determined from a Poincar\'e-covariant Bethe-Salpeter equation (BSE) \cite{Salpeter:1951sz, Nakanishi:1969ph, Lucha:1991vn, Roberts:2015lja, Eichmann:2016yit, Mezrag:2020iuo, Qin:2020rad}.  Its inhomogeneous form may be written:
\begin{equation}
\label{EqBSE}
        \Gamma^H_{\alpha\beta}(k,P)= g^H_{\alpha\beta} + \int_{dq} K^{(2)}_{\alpha\alpha',\beta'\beta} \chi^H_{\alpha'\beta'}(q,P),
\end{equation}
where
$P$ is the total momentum of the quark+antiquark system;
the Bethe-Salpeter wave function is $\chi^H(q,P) = S(q_+) \Gamma^H(q,P) S(q_-)$, with $S(q)$ being the dressed-quark propagator, $q_+ = q+\eta P$, $q_- = q - (1-\eta)P$;
$g^H$ is a combination of Dirac matrices chosen to specify the $J^{PC}$ channel;
$K^{(2)}$ is the two-particle irreducible quark-antiquark scattering kernel, which carries Dirac indices for each of the four fermion legs;
and $\int_{dq}$ denotes a four dimensional Euclidean integral.
%
Herein we consider two flavours of degenerate light-quarks, and suppress all renormalisation constants and color indices for notational simplicity.

The dressed-quark propagator in Eq.\,\eqref{EqBSE} can be computed using the gap equation ($l=k-q$):
\begin{subequations}
\label{EqGap}
\begin{align}
S^{-1}(k) &= i\gamma\cdot k + m + \Sigma(k) \,,\\
\Sigma(k) & =\int_{dq}
4 \pi \alpha\, D_{\mu\nu}(l)\gamma_\mu S(q) \Gamma_\nu(q,k)\,,
\end{align}
\end{subequations}
where $m$ is the Higgs-produced quark current-mass; $\alpha$ is the QCD coupling; $D_{\mu\nu}$ is the dressed-gluon propagator; and $\Gamma_\nu$ is the dressed-gluon-quark vertex.  The solution of Eq.\,\eqref{EqGap} is typically written
$S(k) = 1/[i\gamma\cdot k\,A(k^2) + B(k^2)]$.

The keys to delivering realistic predictions for the meson spectrum lie in beginning with a gap equation kernel that expresses DCSB and therefrom constructing a Bethe-Salpeter kernel which ensures all symmetry constraints germane to the spectrum are preserved \cite{Chang:2009zb, Fischer:2009jm, Chang:2011ei, Williams:2015cvx, Binosi:2016rxz}.
With this goal in mind, consider $K^{(2)}$.  Since this kernel carries four Dirac indices and connects two incoming fermion lines to two outgoing lines, it can be expressed as the sum of tensor products of two $4\times4$ Dirac matrices:
{\allowdisplaybreaks
\begin{subequations}
\begin{align}
\label{eq:kernel}
\rule{-1ex}{0ex}	{K}^{(2)}(q_\pm,k_\pm) & = \sum_n K_{L\,\alpha\alpha^\prime} ^{(n)}(q_\pm,k_\pm) K_{R\,\beta^\prime\beta}^{(n)}(q_\pm,k_\pm) \\
& =: \sum_n K_L^{(n)}(q_\pm,k_\pm) \otimes K_R^{(n)}(q_\pm,k_\pm)\,.
\end{align}
\end{subequations}}
\hspace*{-0.6\parindent}Each $K^{(n)}_{L/R}$ depends on four fermion momenta, only three of which are independent (momentum conservation).
Equation~\eqref{eq:kernel} is general; and with ${K}^{(2)}$ written in this form, it is not necessary that any element $K^{(n)}_{L/R}$ be computable as the sum of a series of diagrams.

For any bound state, $g^H$ in Eq.\,\eqref{EqBSE} specifies the $J^{PC}$ quantum numbers.   ${K}^{(2)}$ is thus even under parity operations, forbidding the following structures Eq.\,\eqref{eq:kernel}:
\begin{equation}
\label{EqParity}
{\mathbf 1} \otimes \gamma_5\,,\;
\gamma_\mu \otimes \gamma_5\gamma_\mu \,,\; {\rm etc.}
\end{equation}
Moreover, ${K}^{(2)}$ is $\mathsf{C}$-parity even, \emph{i.e}.\ with $C$ the charge conjugation matrix, the following identity is required:
\begin{align}
\label{EqCparity}
	{K}^{(2)}(q_\pm,k_\pm) & =
%
 \sum_n {C} [K_R^{(n)}(-k_\mp,-q_\mp)]^{\rm T}{C}^{\dagger} \nonumber \\
&   \qquad \otimes \; {C}[K_L^{(n)}(-k_\mp,-q_\mp)]^{\rm T}{C}^{\dagger}\,.
\end{align}
Finally, since all constructions should respect the connection to a Poincar\'e-invariant local quantum field theory, the $\mathsf{CPT}$ theorem entails that ${K}^{(2)}$ is $\mathsf{T}$-even.  (\emph{N.B}.\ Crossed channels are readily treated.  One merely begins with an appropriately modified form for $g^H$ in Eq.\,\eqref{EqBSE}.)

QCD also has many continuous symmetries, prominent amongst which are those expressed in the vector and axial-vector Ward-Green-Takahashi (WGT) identities \cite{Qin:2013mta, Qin:2014vya}, written compactly here using $\Delta^\pm_F(k)=F(k_-)-F(k_+)$ and $\Delta^\pm_{F5}(k)=F(k_+)\gamma_5 + \gamma_5 F(k_-)$:
\begin{subequations}
\label{eq:WTI}
\begin{align}
	P_\mu \chi_{\mu}(k,P) &= i\Delta_S^\pm(k) \,, \\
        P_\mu \chi_{5\mu}(k,P) &=  \Delta_{S5}^\pm(k)
        - 2im \chi_{5}(k,P) \,.
\end{align}
\end{subequations}
Now using the associated BSEs, defined by $g^H=i\gamma_\mu$, $\gamma_5\gamma_\mu$, and Eq.\,\eqref{EqGap}, Eqs.\,\eqref{eq:WTI} entail
\begin{subequations}
\label{eq:kernelWT}
\begin{align}
	\Sigma(k_+) - \Sigma(k_-)
	& = \sum_n\int_{dq} K_L^{(n)} \Delta^\pm_S(q) K_R^{(n)}\,,\\
	\Sigma(k_+)\gamma_5 + \gamma_5 \Sigma(k_-) & =
	 \sum_n\int_{dq} K_L^{(n)} \Delta^\pm_{S5}(q) K_R^{(n)}\,.
\end{align}
\end{subequations}

In order to resolve Eqs.\,\eqref{eq:kernelWT}, we first decompose
%
\begin{equation}
\label{eq:self5}
\Sigma(k) =  \Sigma_A(k) + \Sigma_B(k)\,,\;
S(k)  = \sigma_A(k) + \sigma_B(k)\,,
\end{equation}
where $\{(\Sigma_A,\sigma_A),\gamma_5\}=0$, $[(\Sigma_B,\sigma_B),\gamma_5]=0$.  Next turning to the Bethe-Salpeter kernel, we write
\begin{align}
{K}^{(2)} & = \left[ K_{L0}^{(+)} \otimes K_{R0}^{(-)} \right] + \left[ K_{L0}^{(-)} \otimes K_{R0}^{(+)} \right] \notag \\
	&\quad + \left[ K_{L1}^{(-)} \otimes_+ K_{R1}^{(-)} \right] + \left[ K_{L1}^{(+)} \otimes_+ K_{R1}^{(+)} \right] \notag \\
	 & \quad + \left[ K_{L2}^{(-)} \otimes_- K_{R2}^{(-)} \right] + \left[ K_{L2}^{(+)} \otimes_- K_{R2}^{(+)} \right]\,,
\label{eq:kernel5}
\end{align}
where $\gamma_5 K^{(\pm)}\gamma_5 = \pm K^{(\pm)}$ and $\otimes_\pm := \tfrac{1}{2}(\otimes \pm \gamma_5 \otimes \gamma_5)$.

Using Eqs.\,\eqref{eq:self5}, \eqref{eq:kernel5}, Eqs.\,\eqref{eq:kernelWT} become
{\allowdisplaybreaks
\begin{subequations}
\label{eq:kernelALL}
\begin{align}
	\Sigma_A&(k_-)  - \Sigma_A(k_+)
= \int_{dq} \Big[ - K_{L0}^{(-)} \sigma_B(q_-) K_{R0}^{(+)} \notag\\
	&+ K_{L0}^{(+)} \sigma_B(q_+) K_{R0}^{(-)}
	 - K_{L2}^{(-)} \Delta_{\sigma_A}^\pm(q) K_{R2}^{(-)} \Big]\,,\label{eq:kernelALL_A}\\
	\Sigma_B&(k_-)  = \int_{dq}\Big[-  K_{L1}^{(-)} \sigma_B(q_-) K_{R1}^{(-)} \notag\\
	&+ K_{L1}^{(+)} \sigma_B(q_+) K_{R1}^{(+)}
    - K_{L0}^{(-)} \Delta_{\sigma_A}^\pm(q) K_{R0}^{(+)} \Big] \,,\label{eq:kernelALL_B}\\
	0 & = \int_{dq} \Big[- K_{L0}^{(-)} \sigma_B(q_+) K_{R0}^{(+)} \notag\\
	&+ K_{L0}^{(+)} \sigma_B(q_-) K_{R0}^{(-)}
	+ K_{L2}^{(+)} \Delta_{\sigma_A}^\pm(q) K_{R2}^{(+)} \Big] \,.\label{eq:kernelALL_A2}
\end{align}
\end{subequations}}
\hspace*{-0.3\parindent}We stress that Eqs.\,\eqref{eq:kernelALL} are simply a decoupled re-expression of the original WGT identities, Eq.\,\eqref{eq:WTI}: no approximation/truncation has been made.  (The path from Eqs.\,\eqref{eq:kernelWT} to \eqref{eq:kernelALL} is detailed in the Supplemental Material.)

\smallskip

\noindent\emph{3:\;Resolving the two-body scattering kernel: functional illustration}.\,---\,%
%
As found when attempting to determine a three-point function from WGT or Slavnov-Taylor identities \cite{Bashir:2011dp, Qin:2013mta, Rojas:2013tza, Qin:2014vya, Aguilar:2014lha, Mitter:2014wpa, Bermudez:2017bpx}, there is no unique solution of the constraint equations~\eqref{EqParity}, \eqref{EqCparity}, \eqref{eq:kernelALL}.  Nevertheless, with a gap equation in hand, one can construct a \emph{minimal} solution for ${K}^{(2)}$ that communicates any emergent features contained in the gap equation kernel to meson properties.  We illustrate this using a gap equation built upon Refs.\,\cite{Qin:2011dd, Chang:2010hb, Qin:2011xq}.

Consider Eq.\,\eqref{EqGap} and set
$4 \pi \alpha D_{\mu\nu}(l) \Gamma_\nu(q,k) \to {\mathpzc G}_{\mu\nu}(l)\Gamma_\nu(q,k)$,
where ${\mathpzc G}_{\mu\nu}$ is a vector-boson exchange-interaction and $\Gamma_\nu$ is a gluon-quark vertex.  A modern form of ${\mathpzc G}_{\mu\nu}(l)$ is explained in Refs.\,\cite{Qin:2011xq, Binosi:2014aea}:
%
${\mathpzc G}_{\mu\nu}(l)  = \tilde{\mathpzc I}(l^2) T_{\mu\nu}(l)$,
with $l^2 T_{\mu\nu}(l) = l^2 \delta_{\mu\nu} - l_\mu l_\nu$ and ($u=l^2$)
\begin{align}
\label{defcalG}
 \tilde{\mathpzc I}(u) & =
 \frac{8\pi^2 D}{\omega^4} e^{-u/\omega^2} + \frac{8\pi^2 \gamma_m \mathcal{F}(u)}{\ln\big[ \tau+(1+u/\Lambda_{\rm QCD}^2)^2 \big]}\,,
\end{align}
where $\gamma_m=4/\beta_0$, $\beta_0=25/3$,
$\Lambda_{\rm QCD}=0.234\,$GeV,
$\ln(\tau+1)=2$,
and ${\cal F}(u) = \{1 - \exp(-u/[4 m_t^2])\}/u$, $m_t=0.5\,$GeV.
Regarding Eq.\,\eqref{defcalG}:
(\emph{i}) $0 < \tilde{\mathpzc I}(0) < \infty$ because a nonzero gluon mass-scale appears as a consequence of EHM in QCD \cite{Binosi:2014aea, Aguilar:2015bud, Cui:2019dwv, Huber:2018ned};
and (\emph{ii}) the large-$u=l^2$ behaviour ensures that the one-loop renormalisation group flow of QCD is preserved.
Quality (\emph{ii}) is important when considering, \emph{e.g}.\ hadron elastic and transition form factors at large momentum transfer \cite{Chen:2018rwz, Ding:2018xwy} and the character of parton distribution functions and amplitudes in the neighbourhood of the endpoints of their support domains \cite{Ding:2018xwy, Ding:2019qlr, Cui:2020dlm, Cui:2020piK}.  However, it plays a far lesser role in the calculation of masses, which are global, integrated properties.  For masses, (\emph{i}) is crucial: even a judiciously formulated momentum-independent interaction can deliver good results \cite{Yin:2019bxe}.  Hence, we follow Refs.\,\cite{Chang:2009zb, Chang:2010hb} and hereafter retain only the first term on the right-hand-side of Eq.\,\eqref{defcalG}.  This simplifies the analysis by obviating renormalisation without materially affecting the results.

The remaining element
is $\Gamma_\nu$. 
In the widely used rainbow-ladder (RL) truncation, $\Gamma_\nu(q,k) = \gamma_\nu$ \cite{Munczek:1994zz, Bender:1996bb}.  For reasons that are understood \cite{Chang:2009zb, Fischer:2009jm, Chang:2011ei, Williams:2015cvx, Binosi:2016rxz}, this \emph{Ansatz} is a good approximation for those bound-states in which orbital angular momentum does not play a significant role and the non-Abelian anomaly can be ignored.  However, it fails for all other systems; its key weakness being omission of those structures which become large as a consequence of EHM.  Such terms typically commute with $\gamma_5$.

The dressed-gluon-quark vertex has twelve independent structures.
In principle, all could be important; but in practice, only five play a material role in the expression of EHM \cite{Binosi:2016wcx}.  Amongst those, the dressed-quark anomalous chromomagnetic moment (ACM) is most important \cite{Skullerud:2003qu, Kizilersu:2006et, Chang:2010hb, Chang:2011ei}: without DCSB, this term vanishes in the chiral limit.  Hence, to illustrate our approach, we use
%
\begin{equation}
\label{EqVertexGap}
\Gamma_\nu(q,k)  = \gamma_\nu + \tau_\nu(l=k-q)\,,\; \tau_\nu(l) = \sigma_{l\nu} \kappa(l^2)\,,
\end{equation}
$\sigma_{l\nu} = \sigma_{\rho\nu} l_\rho$, $\kappa(l^2) = (\eta/\omega)\exp{(-l^2/\omega^2)}$.
$\kappa(l^2)$ is power-law suppressed in QCD; but the Gaussian form, matching the infrared-dominant term in Eq.\,\eqref{defcalG}, is sufficient for illustrative purposes.
In using Eq.\,\eqref{EqVertexGap}, following RL truncation convention, any overall dressing factor $F_1$, as in $F_1(l^2)[\gamma_\nu + \tau_\nu(l)]$, is implicitly absorbed into $\tilde{\mathpzc I}(l^2)$.
%

Returning to the gap equation, Eq.\,\eqref{EqGap}, and introducing the ACM-improved vertex, one can write
\begin{align}
	\Sigma_{A,B}(k_\pm) &= \int_{dq} \mathcal{G}_{\mu\nu}(l) \nonumber \\
& \quad \times \gamma_\mu \left[\sigma_{A,B}(q_\pm) \gamma_\nu + \sigma_{B,A}(q_\pm) \tau_\nu(l) \right] \,.
\end{align}
Using these expressions in Eq.\,\eqref{eq:kernelALL_A}, one obtains
\begin{align}
	{K}^{(2)} &= - \mathcal{G}_{\mu\nu}(l)\gamma_\mu\otimes\gamma_\nu  - \mathcal{G}_{\mu\nu}(l)\gamma_\mu \otimes \tau_\nu(l) \notag\\
	& + ~ \mathcal{G}_{\mu\nu}(l) \tau_\nu(l) \otimes \gamma_\mu +  {K}_{\rm ad} \,.
\end{align}

${K}_{\rm ad}$ is unconstrained by Eq.\,\eqref{eq:kernelALL_A}.  According to Eq.\,\eqref{eq:kernel5}, it only involves $K^{(\mp)}_{L1/R1}$, $K^{(+)}_{L2/R2}$.  To construct a minimal symmetry-consistent kernel, we choose the simplest allowable basis for ${K}_{\rm ad}$.  Given Eqs.\,\eqref{EqVertexGap}, this means
\begin{align}
	{K}_{\rm ad} &= [ \mathbf{1} \otimes_+ \mathbf{1} ] f^{(+)}_{p0} + [ -\mathcal{G}_{\mu\nu}(l)\gamma_\mu \otimes_+ \gamma_\nu ] f^{(-)}_{p1}  \notag\\
	& \quad + [ \mathbf{1} \otimes_- \mathbf{1} ] f^{(+)}_{n0} + [ -\mathcal{G}_{\mu\nu}(l)\sigma_{l\mu} \otimes_- \sigma_{l\nu} ] f^{(+)}_{n1}\,,
	\label{eq:kernel_Ad}
\end{align}
where $f=f(l^2;P^2)\in \mathbb{R}$ for $\{l^2,P^2\}\in \mathbb{R}$.

Inserting Eq.\,\eqref{eq:kernel_Ad} into Eqs.\,\eqref{eq:kernelALL_B}, \eqref{eq:kernelALL_A2}, one obtains:
{\allowdisplaybreaks
\begin{subequations}
\begin{align}
	&\int_{dq} \mathcal{G}_{\mu\nu}(l)\gamma_\mu \sigma_A(q_+) \tau_\nu(l) \notag\\
	&= \int_{dq} \left[ \sigma_B(q_+) f^{(+)}_{p0} + \mathcal{G}_{\mu\nu}(l)\gamma_\mu \sigma_B(q_-) \gamma_\nu f^{(-)}_{p1} \right] \,,\\
	&\int_{dq} \mathcal{G}_{\mu\nu}(l)\gamma_\mu \sigma_B(q_+) \tau_\nu(l) \notag\\
	&= \int_{dq} \left[\sigma_A(q_+) f^{(+)}_{n0} - \mathcal{G}_{\mu\nu}(l)\sigma_{l\mu} \sigma_A(q_+) \sigma_{l\nu} f^{(+)}_{n1} \right] \,.
\end{align}
\end{subequations}}
\hspace*{-0.4\parindent}This pair of complex-valued integral equations gives four real-valued equations that can be solved for the scalar functions which complete $K_{\rm ad}$ and hence $K^{(2)}$.  (See Supplemental Material for exemplifying solutions.)

For an arbitrary vertex in the family specified by Eqs.\,\eqref{EqVertexGap}, we have now arrived at a Bethe-Salpeter kernel that satisfies all necessary and associated discrete and continuous spectrum-generating symmetries.


\smallskip

\noindent\emph{4:\;Impacts of an ACM on the meson spectrum}.\,---\,
%
%
The gap equation's kernel is specified by three parameters: interaction strength, $D$, and range, $\omega$; and ACM strength, $\eta$.  We fix $\omega = 0.8\,$GeV, the value associated with an interaction that matches results from analyses of QCD's gauge sector \cite{Binosi:2014aea, Cui:2019dwv}.  On the other hand, we use $D$ and $\eta$ to highlight the impact of corrections to RL truncation.

First, to establish natural scales, we note that with $D=D_{\rm RL}=(1.105\,{\rm GeV})^2$, $\eta=0$, \emph{i.e}.\ in RL truncation, and with $m=3\,$MeV, the coupled gap and Bethe-Salpeter equations yield $m_\pi =0.14\,$GeV and, using the standard expression \cite{Maris:1997hd}, $f_\pi=0.095\,$GeV.  Both values compare well with empirical results \cite{Zyla:2020}.
Increasing $\eta$ adds DCSB strength to the gap equation's kernel; hence, $D$ must be decreased to maintain the same level of DCSB.  The pairing $D=(0.92\,{\rm GeV})^2$, $\eta=2/5$ yields $f_\pi=0.095\,$GeV, $m_\pi=0.14\,$GeV at $m=3\,$MeV.

It is readily confirmed numerically that our kernel construction preserves both the Gell-Mann--Oakes--Renner relation for meson masses \cite{GellMann:1968rz, Maris:1997hd} and the quark-level Goldberger-Treiman relation \cite{Maris:1997hd, Qin:2014vya}, both of which are salient corollaries of DCSB.  (See Supplemental Material.)

\begin{figure}[t]
\leftline{\hspace*{0.5em}{\large{\textsf{A}}}}
\vspace*{-5ex}
\includegraphics[width=0.84\linewidth]{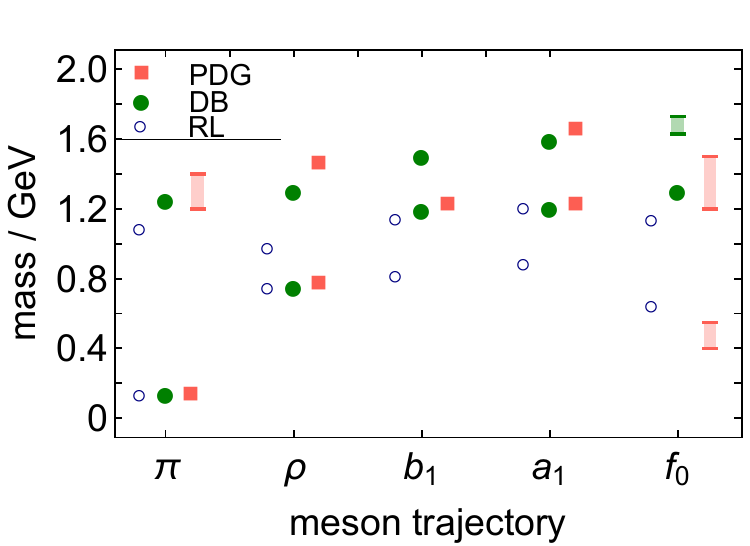}\\[3ex]
\leftline{\hspace*{0.5em}{\large{\textsf{B}}}}
\vspace*{-5ex}
\includegraphics[width=0.84\linewidth]{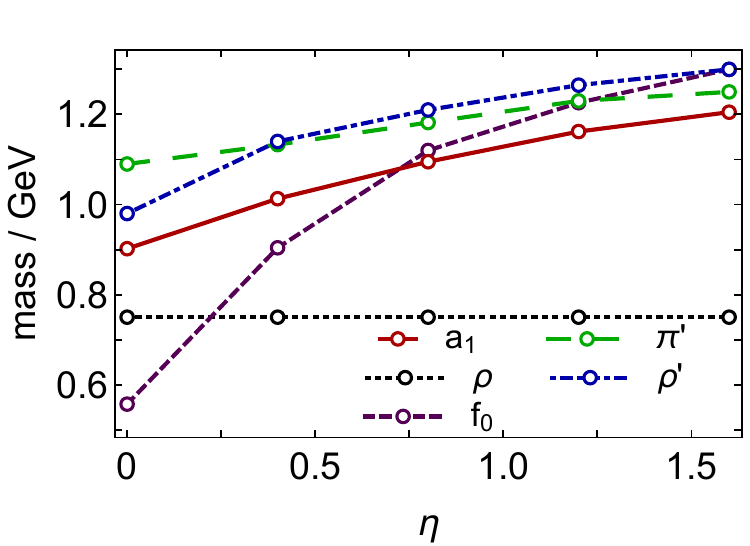}
\caption{\label{fig:spec}
\textsf{A}.
Meson spectrum computed in RL truncation, $D=(1.10\,{\rm GeV})^2$; and using the ACM-corrected dressed-gluon-quark vertex and Bethe-Salpeter kernel (DB, meaning DCSB-improved), $D=(0.72\,{\rm GeV})^2$, $\eta=1.6$.
For comparison, empirical results \cite{Zyla:2020} are also shown, indicated as PDG.  Where bands are drawn, they indicate the quoted mass range.
\textsf{B}.
$\eta$-dependence of selected meson masses, using Eq.\,\eqref{Deta}.
%
In both panels, masses calculated with $m=3\,$MeV.
}
\end{figure}

We now display the impact of the dressed-quark ACM in Eqs.\,\eqref{EqVertexGap} on the meson spectrum.  (The standard BSE solution method is recapitulated in the Supplemental Material.)
First note the computed RL ($\eta=0$) spectrum, represented by the open blue circles in Fig.\,\ref{fig:spec}A.  
As shown by the comparison with empirical values (PDG \cite{Zyla:2020}), the RL masses are too light in almost all cases except the ground-state ($n=1$) $\pi$- and $\rho$-mesons.  The exception is the $f_0$ channel, which is a special case, discussed further below.
All mismatches have long been understood as a systematic flaw of the RL truncation \cite{Roberts:1996jx, Krassnigg:2009zh, Krassnigg:2010mh, Bashir:2012fs, Xu:2019sns}.  Namely, by preserving the vector and axial-vector WGT identities, destructive interferences are ensured between RL correction terms in the ground-state flavour-nonsinglet-pseudoscalar- and vector-channels.  In all other channels the cancellation between corrections is less effective and/or some of the interference between terms is constructive, \emph{i.e}.\ it amplifies pieces of the Bethe-Salpeter kernel that are too weak in RL truncation.

Next, we employ the kernel construction procedure to trace changes in the meson spectrum generated by the ACM term in the gluon-quark vertex.  As $\eta$ is increased from zero, we reduce $D$ so as to keep the $\rho$-meson ground-state mass fixed at $0.75\,$GeV with $m=3\,$MeV:
\begin{equation}
\label{Deta}
D(\eta) = D_{\rm RL}\, (1+0.290 \,\eta)/(1+1.522\,\eta)\,.
\end{equation}
%
The $\eta$-dependence of selected meson masses is depicted in Fig.\,\ref{fig:spec}B.  As found previously with ground-state light-quark mesons \cite{Chang:2011ei}, the EHM-induced ACM term produces considerable improvements over RL truncation.  The following outcomes are worth highlighting.
%
%

(\emph{i}) With increasing $\eta$, the $a_1-\rho$ mass-splitting rises rapidly from the RL result, which is just $1/3$ of the measured value.  The empirical value, $m_{a_1}-m_\rho\approx 0.45\,$GeV, is reproduced at $\eta=1.6$.  (This is the natural size \cite{Chang:2010hb, Binosi:2016wcx}.) Given that current-algebra and related models also only produce $2/3$ of the empirical splitting \cite{Weinberg:1967kj}, this is a significant dynamical outcome with important implications for understanding the meson spectrum.
    For instance, in quantum field theory, one sees the effect as a splitting between parity partners being driven wider by inclusion of additional aspects of DCSB in the gluon-quark coupling.
    Alternatively, from a quark model perspective, which sees the $a_1$ as a $L=1$ quark+antiquark system, it is natural to expect that DCSB-enhanced constituent magnetic moments would increase spin-orbit repulsion, driving the $a_1$ away from the $\rho(L=0)$.
%

(\emph{ii}) The computed mass of the $f_0$ system increases quickly with $\eta$, reaching a value of $\approx 1.3\,$GeV at $\eta=1.6$.  The kernels discussed herein produce a hadron's dressed-quark core.  They do not include the resonant contributions which would typically be associated with a meson-cloud.  This is important because the lightest scalar meson is now considered to be a complicated system with a material $\pi^+\pi^-$ component \cite[Sec.\,62]{Zyla:2020}.  Hence, the quark-core mass of the $f_0$ must be greater than the empirical value because inclusion of resonant contributions to the kernels of the gap and Bethe-Salpeter equations generates additional attraction and a large $f_0\to \pi\pi$  decay width \cite{Holl:2005st, Eichmann:2015cra, Santowsky:2020pwd}.  It is thus notable that the ACM-improved vertex result $m_{f_0} \approx 1.3\,$GeV matches an estimate of the mass of the $q\bar q$-core component of the $f_0$ obtained using unitarised chiral perturbation theory \cite{Pelaez:2006nj}.
%

(\emph{iii}) In RL truncation, the radial excitations of the $\pi$- and $\rho$-mesons are too light and, with $m_{\pi^\prime}-m_{\rho^\prime}\gtrsim 100\,$MeV, ordered incorrectly.  Fig.\,\ref{fig:spec}B shows that both defects can be corrected by including an ACM in the kernels of the gap and Bethe-Salpeter equations.  In fact, at the same value of $\eta=1.6$ that reproduces the empirical value of $m_{a_1}-m_\rho$, $m_{\rho^\prime}-m_{\pi^\prime} \approx 50\,$MeV, commensurate with the empirical value $170(100)\,$MeV.
Given the experimental uncertainty in the $\pi(1300)$ mass, one might doubt that Nature prefers $m_\rho^\prime > m_{\pi^\prime}$; but, in heavy+heavy meson spectra, radial excitations of vector mesons are always slightly heavier than their pseudoscalar partners.
%

We complete Fig.\,\ref{fig:spec}A by including predictions for meson masses obtained using the ACM-corrected gluon-quark vertex specified by Eqs.\,\eqref{EqVertexGap} in formulating the kernels of the gap and Bethe-Salpeter equations.  In addition to the observations already made, our analysis predicts a radial excitation of the $b_1$-meson with mass $m_{b_1^\prime} \approx 1.5\,$GeV.  Such a $2\,^1\!P_1$ state has not yet been seen.

\smallskip

\noindent\emph{5:\;Summary and Perspective}.\,---\,
%
We presented a novel, flexible method for deriving a Bethe-Salpeter kernel for the meson bound-state problem that is symmetry-consistent with any admissible form for the gluon-quark vertex, $\Gamma$.  The construction is applicable even if the diagrammatic content of $\Gamma$ is unknown, as would be the case if the vertex were obtained using lattice-QCD.  It therefore establishes a route to new synergies between continuum and lattice approaches to strong interactions.

The kernel is minimal in the same sense as a resolution of the Ward-Green-Takahashi identity for the photon-quark vertex, $\Gamma^\gamma$: 
it is not the complete result; but it is both a key part of the kernel and a tool enabling demonstrations of consequences of emergent hadronic mass (EHM) that would otherwise be impossible.

The scheme was illustrated using a gluon-quark vertex that includes the EHM-induced dressed-quark anomalous magnetic moment, $\kappa$.  Using a strength for $\kappa$ commensurate with independent estimates, its presence in the vertex and expression in the Bethe-Salpeter kernel were shown to remedy known failings of the commonly used rainbow-ladder (RL) truncation, \emph{e.g}.\ correcting both the mass-splitting between the $a_1$- and $\rho$-mesons and the level ordering of the $\pi$- and $\rho$ meson radial excitations.

As this was the first demonstration of the new scheme, a simplified treatment of quark-antiquark scattering was used.  
Thus, it would be natural to repeat this study using a more realistic interaction.
Moreover, only light-quark mesons were considered.  The spectrum of states in $3$-flavour QCD is much richer, presenting more opportunities for discoveries and increased understanding; and this challenge should be tackled.
Finally, the treatment of baryons using a three valence-quark Faddeev equation is today only possible using the RL truncation.  The scheme described herein can be extended to overcome that limitation.
These efforts are underway.

\smallskip

\noindent\emph{Acknowledgements}.
We are grateful for constructive comments from Z.-F.~Cui and Y.-X.~Liu.
This work was partially completed under the auspices of a Maria Goeppert-Mayer Postdoctoral Fellowship at Argonne National Laboratory.
Work supported by:
National Natural Science Foundation of China under Contracts Nos.\,11805024 and 11947406.



\medskip

\renewcommand{\theequation}{S.\arabic{equation}}
\renewcommand{\thefigure}{S.\arabic{figure}}
\setcounter{equation}{0}
\setcounter{figure}{0}

\centerline{---\,\emph{Supplemental Material}.\,---}

\smallskip

\noindent\emph{Path from Eqs.\,(7) to (10)}.\,---\,
Using Eqs.\,(8), Eqs.\,(7) can be expanded as follows:
\begin{subequations}
\label{BeginSM}
\begin{align}
	& [\Sigma_A(k_+)-\Sigma_A(k_-)] + [\Sigma_B(k_+)-\Sigma_B(k_-)] \nonumber \\
& = \sum_n\int_{dq} K_L^{(n)} [ S(q_-)-S(q_+) ] K_R^{(n)}\,,\\
&	-\gamma_5[\Sigma_A(k_+)-\Sigma_A(k_-)] + \gamma_5[\Sigma_B(k_+)+\Sigma_B(k_-)] \nonumber \\
& = \sum_n\int_{dq} K_L^{(n)} [ S(q_+)\gamma_5 + \gamma_5 S(q_-) ] K_R^{(n)}\,. \label{SMG5}
\end{align}
\end{subequations}

Multiply Eq.\,\eqref{SMG5} from the left with $\gamma_5$ and expand the right-hand-sides of both entries in Eqs.\,\eqref{BeginSM} to obtain:
\begin{subequations}
\label{BeginSM2A}
\begin{align}
&	[\Sigma_A(k_+)-\Sigma_A(k_-)] + [\Sigma_B(k_+)-\Sigma_B(k_-)] \nonumber \\
&= \sum_n\int_{dq} K_L^{(n)} [ \Delta_{\sigma_A}^\pm + \sigma_B(q_-)-\sigma_B(q_+) ] K_R^{(n)}\,,\\
&  -[\Sigma_A(k_+)-\Sigma_A(k_-)] + [\Sigma_B(k_+)+\Sigma_B(k_-)] \nonumber \\
& =\sum_n\int_{dq} \gamma_5 K_L^{(n)} \gamma_5 [ \Delta_{\sigma_A}^\pm + \sigma_B(q_-) + \sigma_B(q_+) ] K_R^{(n)}\,. \label{BeginSM22}
\end{align}
\end{subequations}

Inserting the decomposition in Eq.\,(9) into both entries in Eq.\,\eqref{BeginSM2A} yields:
{\allowdisplaybreaks
\begin{subequations}
\label{BeginSM2}
\begin{align}
[&\Sigma_A (k_+)-\Sigma_A(k_-)] + [\Sigma_B(k_+)-\Sigma_B(k_-)] \nonumber \\
& = \int_{dq} \left\{ K_{L0}^{(+)} [ \Delta_{\sigma_A}^\pm + \sigma_B(q_-)-\sigma_B(q_+) ] K_{R0}^{(-)} \right. \nonumber \\
& \quad + K_{L0}^{(-)} [ \Delta_{\sigma_A}^\pm + \sigma_B(q_-)-\sigma_B(q_+) ] K_{R0}^{(+)} \nonumber \\
&\quad+ K_{L1}^{(-)} [ \sigma_B(q_-)-\sigma_B(q_+) ] K_{R1}^{(-)} \nonumber \\
& \quad+  K_{L1}^{(+)} [ \sigma_B(q_-)-\sigma_B(q_+) ] K_{R1}^{(+)} \nonumber\\
& \quad+ \left. K_{L2}^{(-)} [ \Delta_{\sigma_A}^\pm ] K_{R2}^{(-)}
 + K_{L2}^{(+)} [ \Delta_{\sigma_A}^\pm ] K_{R2}^{(+)} \right\}\,,\\
-[&\Sigma_A(k_+)-\Sigma_A(k_-)] + [\Sigma_B(k_+)+\Sigma_B(k_-)]  \nonumber \\
& =  \int_{dq} \left\{K_{L0}^{(+)} [ \Delta_{\sigma_A}^\pm + \sigma_B(q_-)+\sigma_B(q_+) ] K_{R0}^{(-)} \right. \nonumber\\
& \quad  - K_{L0}^{(-)} [ \Delta_{\sigma_A}^\pm + \sigma_B(q_-)+\sigma_B(q_+) ] K_{R0}^{(+)} \nonumber\\
& \quad- K_{L1}^{(-)} [ \sigma_B(q_-)+\sigma_B(q_+) ] K_{R1}^{(-)} \nonumber\\
&\quad + K_{L1}^{(+)} [ \sigma_B(q_-)+\sigma_B(q_+) ] K_{R1}^{(+)} \nonumber\\
&\quad - \left. K_{L2}^{(-)} [ \Delta_{\sigma_A}^\pm ] K_{R2}^{(-)}
  + K_{L2}^{(+)} [ \Delta_{\sigma_A}^\pm ] K_{R2}^{(+)} \right\}\,,
\end{align}
\end{subequations}}
\hspace*{-0.5\parindent}%
where we have exploited the notation and characteristics described after Eq.\,(9) to arrive at these expressions, \emph{viz}.\,
$\otimes_\pm := \tfrac{1}{2}(\otimes \pm \gamma_5 \otimes \gamma_5)$ and $\gamma_5 K^{(\pm)}\gamma_5 = \pm K^{(\pm)}$.

One can now derive the following relations from appropriate manipulations of Eqs.\,\eqref{BeginSM2}:
{\allowdisplaybreaks
\begin{subequations}
\label{BeginSM4}
\begin{align}
 &\Sigma_B(k_+) = \int_{dq} \left\{ K_{L0}^{(+)} [ \Delta_{\sigma_A}^\pm + \sigma_B(q_-) ] K_{R0}^{(-)} \right. \nonumber\\
	& + K_{L0}^{(-)} [ -\sigma_B(q_+) ] K_{R0}^{(+)}
        +  K_{L1}^{(-)} [ -\sigma_B(q_+) ] K_{R1}^{(-)} \nonumber\\
	& \left. + K_{L1}^{(+)} [ \sigma_B(q_-) ] K_{R1}^{(+)}
        + K_{L2}^{(+)} [ \Delta_{\sigma_A}^\pm ] K_{R2}^{(+)}\right\} \,,\\
[&\Sigma_A(k_+)-\Sigma_A(k_-)]  - \Sigma_B(k_-) \nonumber \\
& = \int_{dq} \left\{ K_{L0}^{(+)} [ -\sigma_B(q_+) ] K_{R0}^{(-)} \right. \nonumber\\
	& +  K_{L0}^{(-)} [ \Delta_{\sigma_A}^\pm + \sigma_B(q_-) ] K_{R0}^{(+)} + K_{L1}^{(-)} [ \sigma_B(q_-) ] K_{R1}^{(-)} \nonumber\\
	& \left. + K_{L1}^{(+)} [ -\sigma_B(q_+) ] K_{R1}^{(+)}+ K_{L2}^{(-)} [ \Delta_{\sigma_A}^\pm ] K_{R2}^{(-)}\right\}\,.
\end{align}
\end{subequations}}

At this point, simple combinations of the form
\begin{equation}
{\rm term}  \pm  \gamma_5\, {\rm term} \, \gamma_5\,,
\end{equation}
where ``term'' is the left- and right-hand-side of a chosen relation in Eq.\,\eqref{BeginSM4}, lead to the following identities:
{\allowdisplaybreaks
\begin{subequations}
\label{BeginSM3}
\begin{align}
	\Sigma_B&(k_+) = \int_{dq} \left\{ K_{L0}^{(+)} [ \Delta_{\sigma_A}^\pm ] K_{R0}^{(-)} -  K_{L1}^{(-)} [ \sigma_B(q_+) ] K_{R1}^{(-)}   \right. \nonumber \\
& \qquad\qquad +\left. K_{L1}^{(+)} [ \sigma_B(q_-) ] K_{R1}^{(+)} \right\}\,,\\
	\qquad 0 &= \int_{dq} \left\{ K_{L0}^{(+)} [ \sigma_B(q_-) ] K_{R0}^{(-)}   -  K_{L0}^{(-)} [ \sigma_B(q_+) ] K_{R0}^{(+)} \right. \nonumber \\
& \qquad\qquad+ \left. K_{L2}^{(+)} [ \Delta_{\sigma_A}^\pm ] K_{R2}^{(+)}\right\}\,,\\
	[\Sigma_A&(k_+)-\Sigma_A(k_-)] = \int_{dq} \left\{  K_{L0}^{(+)} [ -\sigma_B(q_+) ] K_{R0}^{(-)} \right. \nonumber \\
&  \qquad \left. + K_{L0}^{(-)} [ \sigma_B(q_-) ] K_{R0}^{(+)}   + K_{L2}^{(-)} [ \Delta_{\sigma_A}^\pm ] K_{R2}^{(-)}\right\}\,, \\
	- \Sigma_B&(k_-) = \int_{dq} \left\{K_{L0}^{(-)} [ \Delta_{\sigma_A}^\pm ] K_{R0}^{(+)}+ K_{L1}^{(-)} [ \sigma_B(q_-) ] K_{R1}^{(-)} \right. \nonumber \\
& \qquad\qquad \left. + K_{L1}^{(+)} [ -\sigma_B(q_+) ] K_{R1}^{(+)}\right\} \,,
\end{align}
\end{subequations}}
\hspace*{-0.4\parindent}%
which are readily recognised as the entries in Eqs.\,(10).  Here, the expression for $\Sigma_B(k_+)$ is related to that for $\Sigma_B(k_-)$ by charge conjugation; hence it contains no additional information and is omitted from the manuscript.

\smallskip

\noindent\emph{Solving Eqs.\,(16) and Illustrating their Solutions}.\,---\,
Equations~(16) are a pair of complex-valued integral equations, which yield four real-valued linear integral equations whose solutions are the scalar functions that complete $K_{\rm ad}$ and hence $K^{(2)}$.  The simplicity of the equations means they may readily be solved by introducing appropriate quadrature rules to replace the integrations and then working with the matrix equations that result.

\begin{figure}[t]
\includegraphics[width=0.83\linewidth]{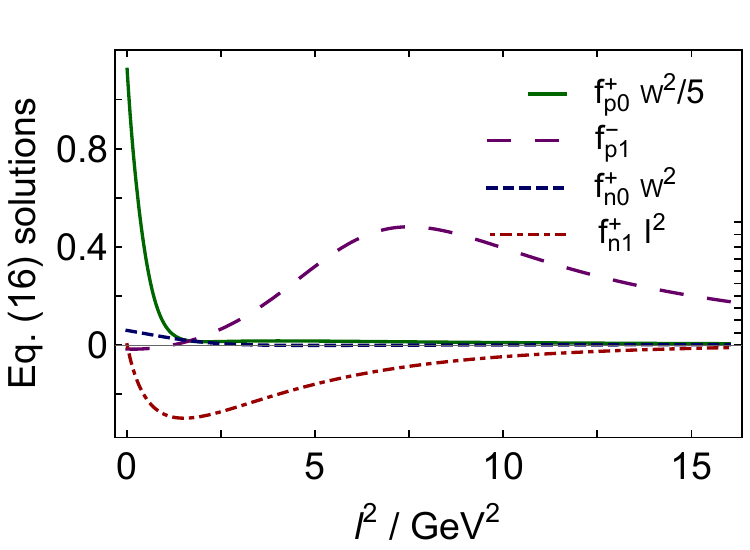}
\caption{
\label{Eq16solutions}
Scalar functions obtained as solutions of Eqs.\,(16), which complete the definition of $K_{\rm ad}$ in Eq.\,(15); hence, $K^{(2)}$.  Each curve depicts a dimensionless function.
Illustration prepared using $D=(0.72\,{\rm GeV})^2$, $\eta=1.6$, $m=3\,$MeV, $P^2=-1\,$GeV$^2$.  Rescaling factor $W^2 = \omega^4/[8\pi^2 D] = (0.1\,{\rm GeV})^2$.
$W^2 f_{p0}^{(+)}$ is overwhelmingly dominant on a material neighbourhood of $l^2=0$, so we divided this function by five in order to present a clear picture.}
\end{figure}

For illustration and future comparisons, it may be useful to draw examples of the solutions obtained.  Using $D=(0.72\,{\rm GeV})^2$, $\eta=1.6$, $m=3\,$MeV, which produce the most favourable results displayed in Fig.\,1, and setting $P^2=-1\,$GeV$^2$, a value roughly midway between that associated with the $\rho$ and $a_1$ mesons, one obtains the results drawn in Fig.\,\ref{Eq16solutions}.   In preparing this image, we referred to Eqs.\,(11), (15) and constructed comparisons of rescaled dimensionless functions: since $1/W^2 = 8\pi^2 D/\omega^4$ is the $l^2=0$ value of ${\mathpzc G}_{\mu\nu}(l)$, it is natural, \emph{e.g}.\ to compare $W^2 f_{p0}^{(+)}$ with $f_{p1}^{(-)}$.

The $f_p$ functions are associated with that part of the WGT identities which relate to the Dirac scalar piece of the quark propagators, whereas the $f_n$ terms are linked with the vector part.   Since DCSB is expressed most strongly in the Dirac scalar part of the propagators, it is therefore unsurprising that the $f_p$ functions are largest.

\smallskip

\begin{figure}[t]
\leftline{\hspace*{0.5em}{\large{\textsf{A}}}}
\vspace*{-5ex}
\includegraphics[width=0.83\linewidth]{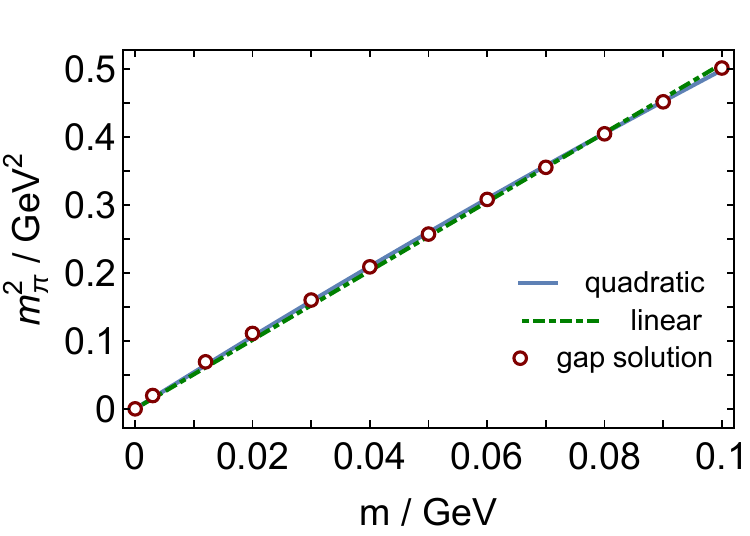}\\[3ex]
\leftline{\hspace*{0.5em}{\large{\textsf{B}}}}
\vspace*{-5ex}
\includegraphics[width=0.85\linewidth]{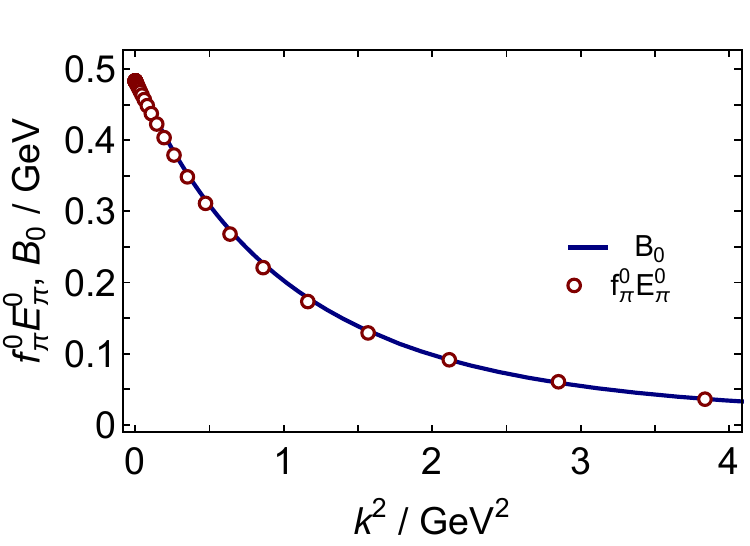}
\caption{\label{fig:GOR}
\emph{Upper panel -- A}.
Pion mass-squared as a function of current-quark mass, $m_\pi^2(m)$: red circles -- numerical solution; solid blue curve -- quadratic fit to numerical result, Eq.\,\eqref{qfit}; and dot-dashed green curve -- linear fit to result, Eq.\,\eqref{lfit}.
\emph{Lower panel -- B}.  Comparison between the two sides of Eq.\,\eqref{EGTR}, \emph{viz}.\ validation of the quark-level Goldberger-Treiman relation.
Results in both panels obtained with $D=(0.92\,{\rm GeV})^2$, $\eta=2/5$.}
\end{figure}

\noindent\emph{GMOR and GT Relations}.\,---\,
Using $D=(0.92\,{\rm GeV})^2$, $\eta=2/5$, we depict $m_\pi^2(m)$ in Fig.\,\ref{fig:GOR}A.  The red circles are the results produced by our Bethe-Salpeter kernel.  They are compared with two fits:
\begin{subequations}
\label{EqFits}
\begin{align}
\label{qfit}
{\rm quadratic:} \quad & m_\pi^2  = m \times 5.40 (1 - 0.077 \, m/m_{\rm m}) \,, \\
{\rm linear:} \quad & m_\pi^2  = m \times 5.07\,,
\label{lfit}
\end{align}
\end{subequations}
where $m_{\rm m}=0.1\,$GeV.  There is little to choose between the fits.  Thus, the kernel we have constructed preserves the Gell-Mann-Oakes-Renner relation \cite{GellMann:1968rz, Maris:1997hd}.

Furthermore, with a computed value of $f_\pi^0=0.093\,$GeV, Eqs.\,\eqref{EqFits} yield the following results for the $m=0$ chiral condensate \cite{Brodsky:2010xf}:
\begin{subequations}
\begin{align}
{\rm quadratic:} \quad & -\langle \bar q q\rangle = (0.286\,{\rm GeV})^3 ,\\
{\rm linear:} \quad & -\langle \bar q q\rangle = (0.280\,{\rm GeV})^3 .
\end{align}
\end{subequations}
They are mutually consistent and compare favourably with typical large renormalisation scale values, \emph{e.g}.\, Refs.\,\cite{Maris:1997tm, Maris:1999nt, Qin:2011dd} find $-\langle \bar q q\rangle\approx (0.276\,{\rm GeV})^3$.

A stringent pointwise test of kernel consistency is provided by the chiral-limit Goldberger-Treiman relation \cite{Maris:1997hd, Qin:2014vya}:
\begin{equation}
\label{EGTR}
f_\pi^0 E_\pi^0(k^2;P^2=0) = B_0(k^2)\,,
\end{equation}
where the index ``0'' indicates that the quantity is calculated in the chiral limit and $E_\pi^0$ is the dominant (pseudoscalar) term in the pion's canonically normalised Bethe-Salpeter amplitude:
\begin{align}
\Gamma_\pi(k;P) &= \gamma_5\left[ i E_\pi(k;P) + \gamma\cdot P F_\pi(k;P) \right. \nonumber\\
& \quad \left. + \gamma\cdot k G_\pi(k;P) + \sigma_{kP} H_\pi(k;P)\right]\,.
\end{align}
Figure\,\ref{fig:GOR}B verifies that our kernel delivers solutions that satisfy this identity.

\smallskip

\noindent\emph{Solving the BSE}.\,---\,
Here we recapitulate a common method for solving the BSE.  Namely, the BSE can be written as an eigenvalue problem: $\Gamma_n = \lambda_n(P^2) K \Gamma_n$.  Here, $\lambda_n(P^2)$ is the eigenvalue; the Bethe-Salpeter amplitude, $\Gamma_n$, is the associated eigenvector; and $\lambda_{n}(P^2)>\lambda_{n+1}(P^2)$ in the absence of level degeneracies.  We use a Euclidean metric, so the on-shell mass for a meson lies at $P^2<0$.

The physical solution for the ground-state, $n=1$, in a given channel is obtained when one finds that timelike value of $P^2$, closest to $P^2=0$, for which $\lambda_0(P^2=-m_{n=1}^2)=1$; the first radial excitation is found by locating the value of $P^2$ for which $\lambda_2(P^2)=1$; etc.\ \cite{Krassnigg:2003wy, Segovia:2015hra}.

Since $P^2<0$ for all physical systems, the variables $q_\pm$, $k_\pm$ in
Eq.\,(3)
are complex valued.
The dressed-quark propagator in the kernel is thus sampled on some domain in the complex plane; and we obtain the solution using now well-known algorithms \cite{Maris:1997tm, Krassnigg:2009gd}.  Those solutions possess complex conjugate poles \cite{Maris:1997tm, Windisch:2016iud}.
With the kernels employed herein, the poles lie outside the sampled domain for meson masses $\lesssim 1.3\,$GeV.  In such cases, the mass and Bethe-Salpeter amplitude are readily obtained.

Today, there are sophisticated methods \cite{Chang:2013nia}
based on perturbation theory integral representations \cite{Nakanishi:1969ph}
for handling states with mass $\gtrsim 1.3\,$GeV.  They provide access to the meson mass and Bethe-Salpeter amplitude.  However, they are cumbersome to implement.
Herein, since we are only interested in masses, we employ the eigenvalue extrapolation procedure introduced in Ref.\,\cite{Frank:1995uk}; to wit, for the heavier systems, we compute $\lambda_n(P^2)$ on a $P^2$-domain that is unaffected by the propagator poles and then extrapolate in $P^2$ to locate the zero of $[1-\lambda_n(P^2)]$.  This yields an estimate of the meson's mass along with an uncertainty.  It does not provide straightforward access to the associated Bethe-Salpeter amplitude.

In preparing Fig.\,1A, eigenvalue extrapolations were used for the $b_1^\prime$, $a_1^\prime$, $f_0^\prime$.  In the first two cases, the uncertainty is smaller than the size of the associated plot marker.  In the last case, it is a little larger; so we display a band that expresses the extrapolation uncertainty.


\end{document}